\documentclass[aps,twocolumn,showpacs,groupedaddress,amsmath,amssymb]{revtex4-1}

\usepackage{graphicx}
\usepackage{color}

\newcommand{\eqreff}[1]{Equation~(\ref{#1})}

\newcommand{\be}{\begin{equation}}
\newcommand{\ee}{\end{equation}}
\newcommand{\bes}{\begin{eqnarray}}
\newcommand{\ees}{\end{eqnarray}}
\newcommand{\bess}{\begin{eqnarray*}}
\newcommand{\eess}{\end{eqnarray*}}

\begin{document}

\title{Fluctuation Theorem of Information Exchange \\between Subsystems that Co-Evolve in Time}

\author{Lee Jinwoo}
\email{e-mail: jinwoolee@kw.ac.kr}
\affiliation{Department of Mathematics, Kwangwoon 
University, 20 Kwangwoon-ro, Nowon-gu, Seoul 139-701, Korea}

\date{\today}

\begin{abstract}
Sagawa and Ueda established a fluctuation theorem of information exchange by revealing the role of correlations in stochastic thermodynamics and unified the non-equilibrium thermodynamics of measurement and feedback control [T. Sagawa and M. Ueda, Phys. Rev. Lett. 109, 180602 (2012)]. They considered a process where a non-equilibrium system exchanges information with other degrees of freedom such as an observer or a feedback controller. They proved the fluctuation theorem of information exchange under the assumption that the state of the other degrees of freedom that exchange information with the system does not change over time while the states of the system evolve in time. Here we relax this constraint and prove that the same form of the fluctuation theorem holds even if both subsystems co-evolve during information exchange processes. This result \textcolor{black}{may extend} the applicability of the fluctuation theorem of information exchange to a broader class of non-equilibrium processes, \textcolor{black}{such as a dynamic coupling in biological systems, where subsystems that exchange information interact with each other.}
\end{abstract}

\maketitle

\section{Introduction}

\textcolor{black}{
Biological systems possess information processing mechanisms for their survival and heredity \cite{hartwell1999molecular,crofts2007life,cheong2011information}. They, for example, sense external ligand concentrations \cite{Thomas2017, Ouldridge2017}, transmit information through signaling networks \cite{Becker2015, cheng2016, Guo2016}, and coordinate gene expressions \cite{olimpio2018} by secreting and sensing signaling molecules \cite{maire2015}. Cells even implement time integration by copying states of environment into molecular states inside the cells to reduce their sensing errors \cite{mehta2012energetic, govern2014energy}. Therefore it is crucial to reveal the role of information in thermodynamics to properly understand complex biological information processes.}

\textcolor{black}{Historically}, information has entered into the realm of thermodynamics by the name of Maxwell's demon. The demon observes the speed of molecules in a box that is divided into two portions by a partition in which there is a small hole, and lets the fast particles pass from the lower-half of the box to the upper-half, and only the slow particles pass from the upper-half to the lower-half by opening/closing the hole without expenditure of work (see Figure 1a). This results in raising the temperature of the upper-half of the box and lower that of the lower-half, indicating that the second law of thermodynamics, which implies heat flows spontaneously from hotter to colder places, might hypothetically be violated \cite{leff2014}. This paradox shows that information can affect thermodynamics of a physical system, or information is a physical element \cite{landauer1991information}.

\begin{figure*}[t]
\centering
\includegraphics[width=15cm]{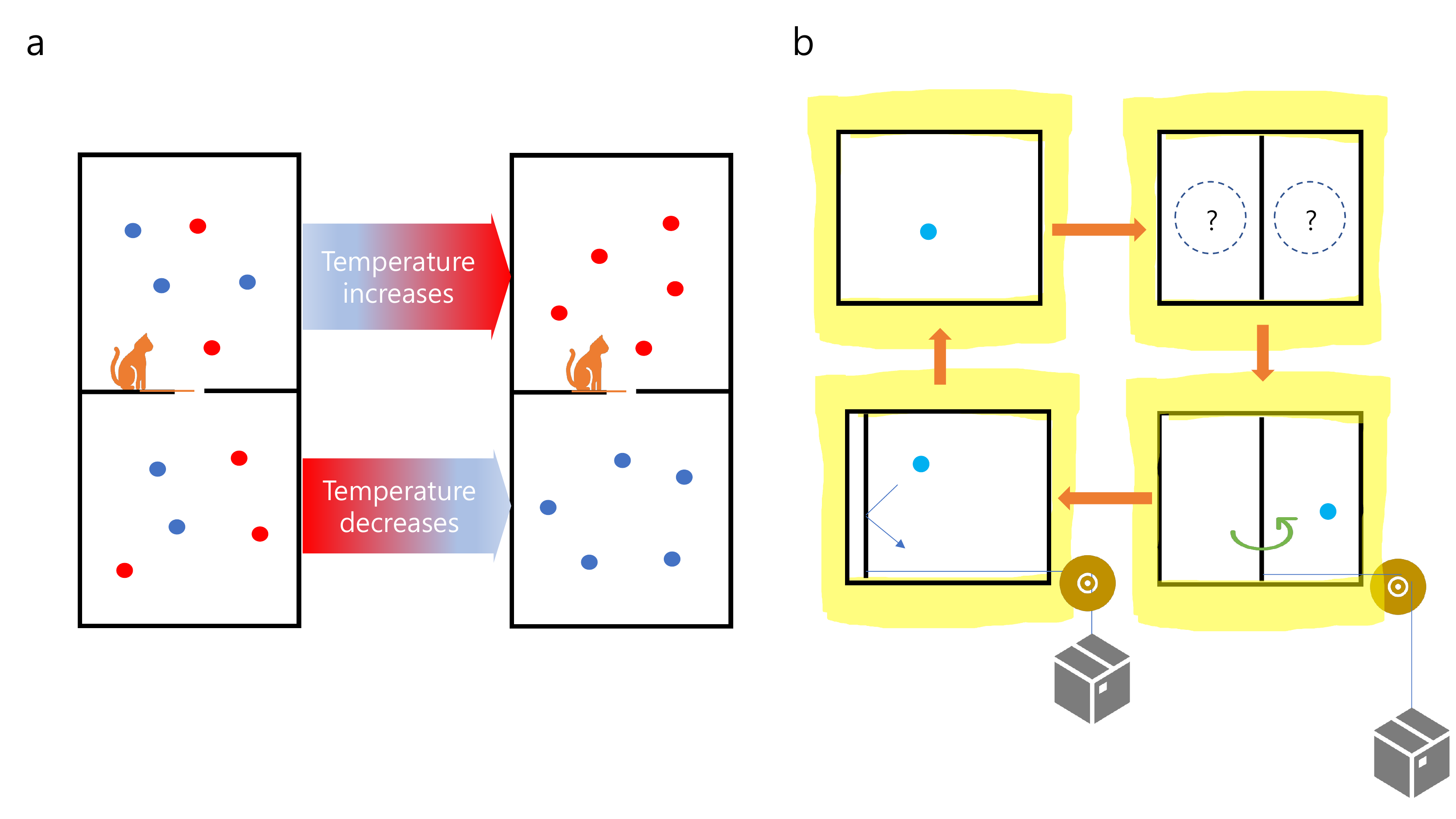}
\caption{Paradox in thermodynamics of information (\textbf{a}) Maxwell's demon (orange cat) uses information on the speed of the particles in the box: He opens/closes the small hole (orange line) without expenditure of energy such that fast particles (red filled circles) are gathered in the upper-half of the box and slow particles (blue filled circles) are gathered in the lower-half of the box. Since temperature is the average velocity of the particles, the demon's action results in spontaneous flow of heat from colder places to hotter places, which violates the second-law of thermodynamics. (\textbf{b}) \textcolor{black}{A cycle of Szilard's engine is represented.}  A lever (green curved arrow) is controlled such that a weight can be lifted during the wall moves quasi-statically in the direction that the particle pushes. \textcolor{black}{This engine harnesses heat from the heat reservoir (yellow region around each boxes) and convert it into mechanical work, cyclically, and thus} corresponds to a perpetual-motion engine of the second kind, which is prohibited by the second-law of thermodynamics.}
\end{figure*}

Szilard has devised a much simpler model that carries the essential role of information in Maxwell's thought experiment. The Szilard engine consists of a single particle in a box which is surrounded by a heat reservoir of constant temperature. A cycle of the engine begins with inserting a partition in the middle of the box. Depending on whether the particle is in the left-half or in the right-half of the box, one controls a lever such that \textcolor{black}{a weight can be lifted during the wall moves quasi-statically in the direction that the particle pushes} (see Figure 1b). If the partition reaches an end of the box, the partition is removed and a new cycle begins again with inserting a partition at the center. Since the energy required for lifting the weight comes from the heat reservoir, this engine corresponds to a perpetual-motion machine of the second kind, where the single heat reservoir is spontaneously cooled and the corresponding thermal energy is converted into mechanical work cyclically, which is prohibited by the second-law of thermodynamics \cite{szilard}.
 
Szilard interprets the coupling between the location of the particle and the direction of the lever as a sort of memory faculty and points out that the coupling is the main cause that enables an amount of work to be extracted from the heat reservoir. He infers,  therefore, that establishing the coupling must be accompanied by a production of entropy (dissipation of heat into the environment) which compensates for the lost heat in the reservoir. In \cite{sagawa2}, Sagawa and Ueda have proved this idea in the form of a fluctuation theorem of information exchange, generalizing the second-law of thermodynamics by taking information into account:
\be\label{eq:1}
\left< e^{-\sigma+\Delta I}\right> = 1,
\ee 
where $\sigma$ is the entropy production of a system $X$, and $\Delta I$ is the change of mutual information between the system $X$ and another system $Y$, \textcolor{black}{such as a demon, during a process $\lambda_t$ for $0\le t\le \tau$.} Here the bracket indicates the ensemble average over all microscopic trajectories \textcolor{black}{of $X$ and over all states of $Y$.} By Jensen's inequality \cite{cover}, \eqreff{eq:1} implies 
\be\label{eq:2}
\left<\sigma\right> \ge \left< \Delta I\right>.
\ee

This tells indeed that establishing a correlation between the two subsystems, $\left<\Delta I \right> > 0$, accompanies an entropy production, $\left<\sigma\right> > 0$, and expenditure of this correlation,  $\left<\Delta I \right> < 0$, serves as a source of entropy decrease, $\left<\sigma\right> < 0$. In proving this theorem, they have assumed that the state of system $Y$ does not evolve in time. \textcolor{black}{This assumption causes no problem for simple models of measurement and feedback control. However, in biological systems, it is not unusual that both subsystems that exchange information with each other co-evolve in time. 
\textcolor{black}{For example, transmembrane receptor proteins transmit signals through thermodynamic coupling between extracellular ligands and conformation of intracellular parts of the receptors during a dynamic allosteric transition \cite{tsai2014unified, cuendet2016allostery}.} In this paper, we relax the constraint that Sagawa and Ueda have assumed, and generalize the fluctuation theorem of information exchange to \textcolor{black}{be applicable to} more involved situations, where the two subsystems can influence each other so that the states of both systems co-evolve in time.} 




\section{Results}
\subsection{Theoretical Framework}
We consider a finite classical stochastic system composed of subsystems $X$ and $Y$ that are in contact with a heat reservoir of inverse temperature $\beta\equiv 1/(k_BT)$ where $k_B$ is the Boltzmann constant and $T$ is the temperature of the reservoir. We allow both systems $X$ and $Y$ to be driven far from equilibrium by changing external parameter $\lambda_t$ during time $0\le t\le \tau$ \cite{jarReview,revSeifert,review}. We assume that time evolutions of subsystems $X$ and $Y$ are described by a classical stochastic dynamics from $t=0$ to $t=\tau$ along trajectories $\{x_t\}$ and $\{y_t\}$, respectively, where $x_t$ ($y_t$) denotes a specific microstate of $X$ ($Y$) at time $t$ for $0\le t\le \tau$ on each trajectory. 
Since both trajectories fluctuate, we repeat the process $\lambda_t$ with appropriate initial joint probability distribution $p_0(x,y)$ over all microstates $(x, y)$ of systems $X$ and $Y$. Then the joint probability distribution $p_t(x,y)$ would evolve for $0\le t\le \tau$. \textcolor{black}{Let $p_t(x):=\int p_t(x,y)\,dy$ and $p_t(y):=\int p_t(x,y)\,dx$ be the corresponding} marginal probability distributions.
We assume
\be\label{eq:neq}
p_0(x,y) \neq 0 \mbox{ for all } (x,y)
\ee
so that we have  
$p_t(x,y) \neq 0$, $p_t(x) \neq 0$, and $p_t(y) \neq 0$ for all $x$ and $y$ during $0\le t \le\tau$.  

Now, the entropy production $\sigma$ during process $\lambda_t$ for $0\le t\le \tau$ is given by
\be\label{eq:sigma}
\sigma:=\Delta s  + \beta Q_b,
\ee 
where $\Delta s$ is the sum of changes in stochastic entropy along $\{x_t\}$ and $\{y_t\}$, and $Q_b$ is heat dissipated into the reservoir (entropy production in the reservoir) \cite{crooks99,seifert05}. In detail, we have
\bes\label{eq:s}
\begin{aligned}
\Delta s &:= \Delta s_x + \Delta s_y,\\
\Delta s_x &:= -\ln p_\tau(x_\tau) + \ln p_0(x_0), \\
\Delta s_y &:= -\ln p_\tau(y_\tau) + \ln p_0(y_0).
\end{aligned}
\ees

We note that the stochastic entropy $s[p_t(\circ)]:=-\ln p_t(\circ)$ of microstate $\circ$ at time $t$ can be interpreted as uncertainty of occurrence of $\circ$ at time $t$: The greater the probability that state $\circ$ occurs, the smaller the uncertainty of occurrence of state $\circ$. 

Now we consider situations where system $X$ exchanges information with system $Y$ during process $\lambda_t$. By this, we mean that trajectory $\{x_t\}$ of system $X$ evolves depending on the trajectory $\{y_t\}$ of system $Y$. Then, information $I_t$ at time $t$ between $x_t$ and $y_t$ is characterized by the \textcolor{black}{reduction of uncertainty of $x_t$ due to given $y_t$} \textcolor{black}{\cite{sagawa2}}: 
\bes\label{eq:I}
\begin{aligned}
I_t(x_t,y_t) &:= s[p_t(x_t)] - s[p_t(x_t | y_t)] \\
      &\,\, = \ln \frac{p_t(x_t, y_t)}{p_t(x_t)p_t(y_t)},
\end{aligned}
\ees
where $p_t(x_t|y_t)$ is the conditional probability distribution of $x_t$ given $y_t$. \textcolor{black}{We note that
this is called the (time-dependent form of)} thermodynamic coupling function \cite{cuendet2016allostery}.
The larger the value of $I_t(x_t,y_t)$ is, the more information \textcolor{black}{is being shared between $x_t$ and $y_t$ for their occurrence.} We note that $I_t(x_t,y_t)$ vanishes if $x_t$ and $y_t$ are independent at time $t$, and the average of $I_t(x_t,y_t)$ with respect to $p_t(x_t,y_t)$ over all microstates is the mutual information between the two subsystems, which is greater than or equal to zero \cite{cover}. 

\subsection{Proof of Fluctuation Theorem of Information Exchange}
Now we are ready to prove the fluctuation theorem of information exchange in this general setup. 
We define reverse process $\lambda'_t := \lambda_{\tau-t}$ for $0\le t\le \tau$, where the external parameter is time-reversed \cite{ponmurugan2010,horowitz2010}. 
Here we set the initial probability distribution $p'_0(x,y)$ for the reverse process as the final (time $t=\tau$) probability distribution for the forward process $p_\tau(x,y)$ so that we have 
\bes\label{eq:pp}
\begin{aligned}
p'_0(x)=\int p'_0(x,y)\, dy = \int p_\tau(x,y)\,dy = p_\tau(x), \\
p'_0(y)=\int p'_0(x,y)\, dx = \int p_\tau(x,y)\,dx = p_\tau(y). 
\end{aligned}
\ees
Then, by \eqreff{eq:neq}, we have $p'_t(x,y) \neq 0$, $p'_t(x) \neq 0$, and $p'_t(y) \neq 0$ for all $x$ and $y$ during $0\le t \le\tau$. 
We also consider the time-reversed conjugate for each $\{x_t\}$ and $\{y_t\}$ for $0\le t\le \tau$ as follows:
\bes
\begin{aligned}
\{x'_t\}:=\{x^*_{\tau-t}\}, \\
\{y'_t\}:=\{y^*_{\tau-t}\},
\end{aligned}
\ees
where $*$ denotes momentum reversal. The microscopic reversibility condition connects the time-reversal symmetry of the microscopic dynamics to non-equilibrium thermodynamics, and reads in this framework as follows \cite{kur, maes1999,crooks99,jar2000}:
\be\label{eq:Qb}
\frac{p(\{x_t\}, \{y_t\} | x_0,y_0)}{p'(\{x'_t\},\{y'_t\} | x'_0,y'_0)}=e^{\beta Q_b},
\ee
where $p(\{x_t\}, \{y_t\} | x_0,y_0)$ is the conditional joint probability distribution of paths $\{x_t\}$ and $\{y_t\}$ conditioned at initial microstates $x_0$ and $y_0$, and $p'(\{x'_t\},\{y'_t\} | x'_0,y'_0)$ is that for the reverse process.   
Now we have the following:
\begin{widetext}
\bes\label{eq:l1}
\frac{p'(\{x'_t\},\{y'_t\} )}{p(\{x_t\},\{y_t\})} & = & 
\frac{p'(\{x'_t\},\{y'_t\} | x'_0,y'_0)}{p(\{x_t\}, \{y_t\} | x_0,y_0)}\cdot \frac{p'_0(x'_0,y'_0)}{p_0(x_0,y_0)} \\  \label{eq:l2}
& = & \frac{p'(\{x'_t\},\{y'_t\} | x'_0,y'_0)}{p(\{x_t\}, \{y_t\} | x_0,y_0)}\cdot\frac{p'_0(x'_0,y'_0)}{p'_0(x'_0)p'_0(y'_0)}\cdot
\frac{p_0(x_0)p_0(y_0)}{p_0(x_0,y_0)}\cdot\frac{p'_0(x'_0)}{p_0(x_0)}\cdot\frac{p'_0(y'_0)}{p_0(y_0)}\\ \label{eq:l3}
& = & \exp\{ - \beta Q_b + I_\tau(x_\tau,y_\tau) - I_0(x_0,y_0) - \Delta s_x - \Delta s_y\}\\ \label{eq:l4}
& = & \exp\{- \sigma + \Delta I \}.
\ees
\end{widetext}

To obtain \eqreff{eq:l2} from \eqreff{eq:l1}, we multiply \eqreff{eq:l1} by $\frac{p'_0(x'_0)p'_0(y'_0)}{p'_0(x'_0)p'_0(y'_0)}$ and $\frac{p_0(x_0)p_0(y_0)}{p_0(x_0)p_0(y_0)}$, which are $1$. 
We obtain \eqreff{eq:l3} by applying Equations \eqref{eq:s}--\eqref{eq:pp} and \eqref{eq:Qb}  consecutively to \eqreff{eq:l2}.
Finally, we set $\Delta I := I_\tau(x_\tau,y_\tau) - I_0(x_0,y_0)$, and use \eqreff{eq:sigma} to obtain \eqreff{eq:l4} from \eqreff{eq:l3}. 

We note that \eqreff{eq:l4} generalizes the detailed fluctuation theorem in the presence of information exchange that is proved in \cite{sagawa2}. Now we obtain the generalized version of \eqreff{eq:1} by using \eqreff{eq:l4} as follows:  
\bes\label{eq:main}
\begin{aligned}
\left<e^{-\sigma+\Delta I}\right> &= \int e^{-\sigma+\Delta I} p(\{x_t\},\{y_t\}) \, d\{x_t\}d\{y_t\} \\
& = \int p'(\{x'_t\},\{y'_t\} ) \, d\{x'_t\}d\{y'_t\} = 1.
\end{aligned}
\ees

Here we use the fact that there is a one-to-one correspondence between the forward and the reverse paths due to the time-reversal symmetry of the underlying microscopic dynamics such that $d\{x_t\}=d\{x'_t\}$ and $d\{y_t\}=d\{y'_t\}$ \cite{goldstein}.

\subsection{Corollary}
Before discussing a corollary, we remark one thing: 
we have used similar notation to that used by Sagawa and Ueda in \cite{sagawa2}, but there is an important difference.
Most importantly, their entropy production $\sigma_{\rm su}$ reads as follows:
\be\nonumber
\sigma_{\rm su} := \Delta s_{\rm su} + \beta Q_b,
\ee
where $\Delta s_{\rm su}:=\Delta s_x$. In \cite{sagawa2}, system $X$ is in contact with the heat reservoir, but system $Y$ is not. Nor does system $Y$ evolve over time. Thus they have considered entropy production in system $X$ and the bath. 
In this paper, both systems $X$ and $Y$ are in contact with the reservoir, and system $Y$ also evolves in time. Thus both subsystems $X$ and $Y$ as well as the heat bath contribute to the entropy production as expressed in Equations \eqref{eq:sigma} and \eqref{eq:s}. Keeping in mind this difference, \textcolor{black}{we apply Jensen's inequality to \eqreff{eq:main} to obtain}  
\be\label{eq:coro}
\left<\sigma\right> \ge \left< \Delta I\right>.
\ee
It tells us that firstly, establishing correlation between $X$ and $Y$ accompanies entropy production, and secondly, established correlation serves as a source of entropy decrease. 

\begin{figure*}[t]
\centering
\includegraphics[width=15cm]{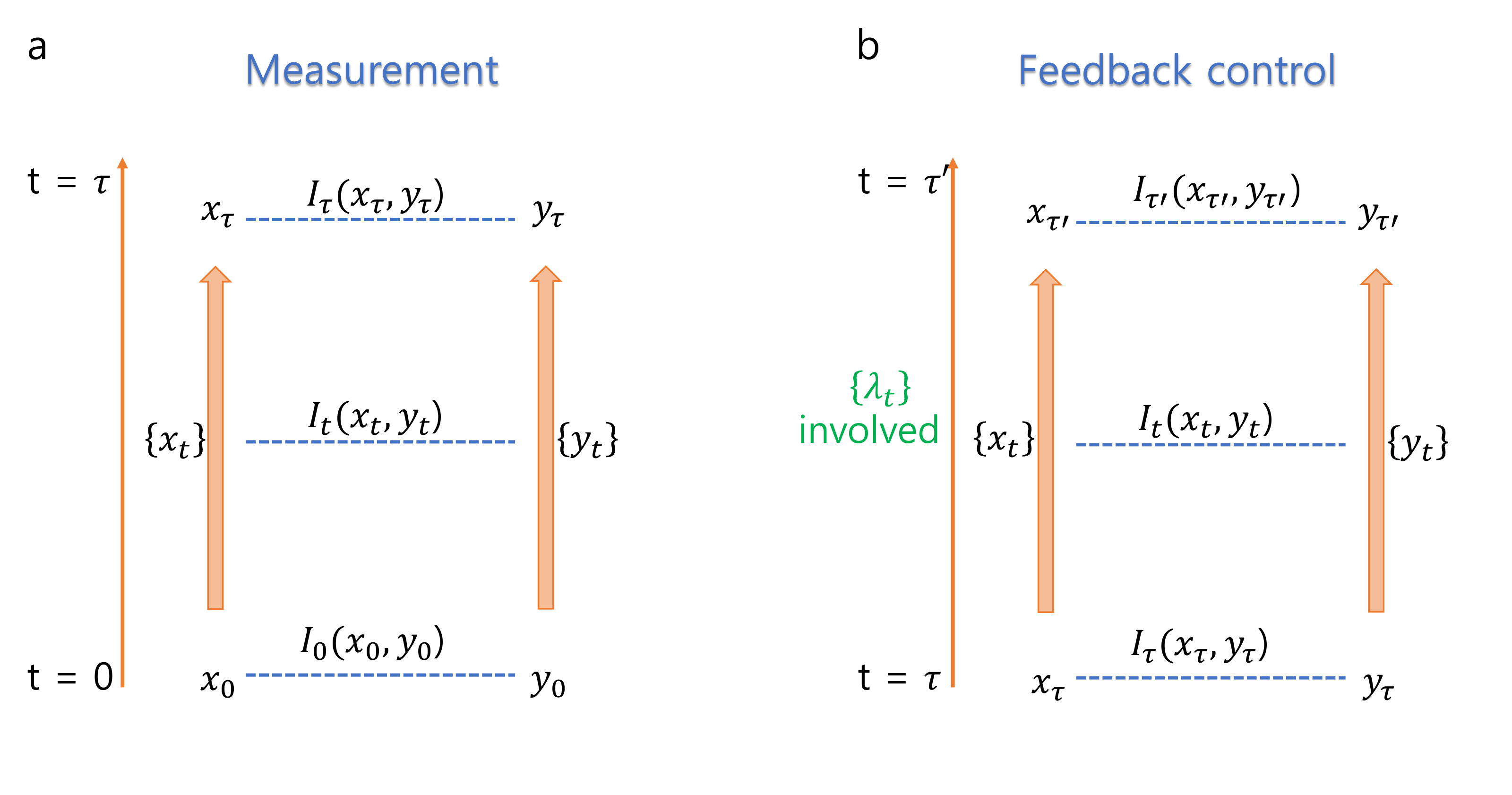}
\caption{
\textcolor{black}{
Measurement and feedback control: system $X$ is, for example, a measuring device and system $Y$ is a measured system.
$X$ and $Y$ co-evolve, as they interact \textcolor{black}{weakly}, along trajectories $\{x_t\}$ and $\{y_t\}$, respectively.
(\textbf{a}) Coupling is being established during the measurement process so that $I_t(x_t,y_t)$ for $0\le t\le \tau$ may be increased (not necessarily monotonically).  
(\textbf{b}) Established correlation is being used as a source of work through external parameter $\lambda_t$ so that $I_t(x_t,y_t)$ for $\tau \le t\le \tau'$ may be decreased (not necessarily monotonically).
}
}
\end{figure*}

\textcolor{black}{Now as a corollary, we refine the generalized fluctuation theorem in \eqreff{eq:main} by including energetic terms.
To this end, we define local free energy ${\cal F}_x$ of system $X$ at $x_t$ and ${\cal F}_y$ of system $Y$ at $y_t$ as follows: 
\bes\label{eq:psi}
\begin{aligned}
{\cal F}_x (x_t,  t) &:= E_x(x_t, t) - Ts[p_t(x_t)]  \\
{\cal F}_y (y_t,  t) &:= E_y(y_t, t) - Ts[p_t(y_t)],
\end{aligned}
\ees
where $E_x$ and $E_y$ are internal energy of systems $X$ and $Y$, respectively, and $s[p_t(\circ)]:=-\ln p_t(\circ)$ is stochastic entropy \textcolor{black}{\cite{crooks99,seifert05}}. Here $T$ is the temperature of the heat bath and argument $t$ indicates dependency of each terms on external parameter $\lambda_t$. During the process $\lambda_t$, work done on the systems is expressed by the first law of thermodynamics as follows:
\be\label{eq:1st}
W := \Delta E + Q_b,
\ee
where $\Delta E$ is the change in internal energy of the systems. 
If we assume that systems $X$ and $Y$ are weakly coupled, in that interaction energy between $X$ and $Y$ is negligible compared to internal energy of $X$ and $Y$, we may have 
\be\label{eq:E}
\Delta E := \Delta E_x + \Delta E_y,
\ee
where $\Delta E_x := E_x(x_\tau,\tau) -E_x(x_0,0)$ and $\Delta E_y := E_y(y_\tau,\tau) -E_y(y_0,0)$ \cite{parrondo2015}.
We rewrite \eqreff{eq:l3} by adding and subtracting the change of internal energy $\Delta E_x$ of $X$ and $\Delta E_y$ of $Y$ as follows:
\begin{widetext}
\bes\label{eq:ll1}
\frac{p'(\{x'_t\},\{y'_t\} )}{p(\{x_t\},\{y_t\})}
& = & \exp\{ - \beta (Q_b + \Delta E_x + \Delta E_y) + \Delta I  + \beta \Delta E_x  - \Delta s_x + \beta \Delta E_y  - \Delta s_y\}\\
\label{eq:ll2}
& = & \exp\{- \beta (W - \Delta {\cal F}_x - \Delta {\cal F}_y) + \Delta I   \},
\ees
\end{widetext}
where we have applied Equations \eqref{eq:psi}--\eqref{eq:E} consecutively to \eqreff{eq:ll1}  to obtain \eqreff{eq:ll2}. 
Here $\Delta {\cal F}_x:={\cal F}_x(x_\tau,\tau)-{\cal F}_x(x_0,0)$ and $\Delta{\cal F}_y := {\cal F}_y(y_\tau,\tau)-{\cal F}_y(y_0,0)$.
Now we obtain fluctuation theorem of information exchange with energetic terms as follows:
\begin{widetext}
\bes\label{eq:main2}
\begin{aligned}
\left<e^{- \beta (W - \Delta {\cal F}_x - \Delta {\cal F}_y) + \Delta I } \right> 
&= \int e^{- \beta (W - \Delta {\cal F}_x - \Delta {\cal F}_y) + \Delta I   }  p(\{x_t\},\{y_t\}) \, d\{x_t\}d\{y_t\} \\
& = \int p'(\{x'_t\},\{y'_t\} ) \, d\{x'_t\}d\{y'_t\} = 1,
\end{aligned}
\ees
\end{widetext}
which generalizes known relations in the literature \cite{kawai, takara, hasegawa, esposito2011, sagawa,parrondo2015}.
We note that \eqreff{eq:main2} holds under the weak-coupling assumption between systems $X$ and $Y$ during the process $\lambda_t$.
By Jensen's inequality, \eqreff{eq:main2} implies 
\be\label{eq:coro2}
\left<W \right> \ge \left< \Delta {\cal F}_x + \Delta {\cal F}_y + \frac{\Delta I}{\beta} \right>.
\ee
\indent We remark that $\left< \Delta {\cal F}_x\right> + \left<\Delta {\cal F}_y\right>$ in \eqreff{eq:coro2} is the difference in non-equilibrium free energy, which is different from the change in equilibrium free energy that appears in similar relations in the literature \cite{kawai, takara, hasegawa, esposito2011, sagawa}.
}

\section{Examples}
\subsection{Measurement}
\textcolor{black}{
Let $X$ be a device (or a demon) which measures the state of other system and $Y$ be a measured system, both of which are in contact with a heat bath of inverse temperature $\beta$ (see Figure 2a). We consider a dynamic measurement process, which is described as follows: $X$ and $Y$ are prepared separately in equilibrium such that $X$ and $Y$ are not correlated initially, i.e.,  $I_0(x_0,y_0)=0$ for all $x_0$ and $y_0$. At time $t=0$, device $X$ is put in contact with system $Y$ so that the coupling of $X$ and $Y$ occurs due to their (weak) interactions until time $t=\tau$, at which a single measurement process finishes. We note that system $Y$ is allowed to evolve in time during the process. Since each process fluctuates, we repeat the measurement many times to obtain probability distribution $p_t(x,y)$ for $0\le t\le \tau$.}

\textcolor{black}{A distinguished feature of the framework in this paper is that mutual information $I_t(x_t, y_t)$ in \eqreff{eq:I} enables us to obtain the time-varying amount of established information during the dynamic coupling process, unlike other approaches where they either provide the amount of information at a fixed time \cite{sagawa, horowitz2011thermodynamic, parrondo2015} or one of the system is fixed during the coupling process \cite{sagawa2}. For example, let us assume that the probability distribution $p_t(x_t, y_t)$ at an intermediate time $t$ is as shown in Table 1.} 

\begin{table}[h]
\caption{The joint probability distribution of $x$ and $y$ at \textcolor{black}{an intermediate time $t$: Here we assume for simplicity that both systems $X$ and $Y$ have two states, $0$ (left) and $1$ (right).}}
\centering
\begin{tabular}{ccc}
\hline
\textbf{${X\backslash Y}$}	& \textbf{0 (Left)}	& \textbf{1 (Right)}\\
\hline
\textbf{0 (Left)}		& 1/3			& 1/6\\
\textbf{1 (Right)}		& 1/6			& 1/3\\
\hline
\end{tabular}
\end{table}

Then we have the following:
\bes\label{eq:I0}
\begin{aligned}
I_t(x_t=0,y_t=0) &= \ln\frac{1/3}{(1/2)\cdot(1/2)} = \ln (4/3), \\
I_t(x_t=0,y_t=1) &= \ln\frac{1/6}{(1/2)\cdot(1/2)} = \ln (2/3), \\
I_t(x_t=1,y_t=0) &= \ln\frac{1/6}{(1/2)\cdot(1/2)} = \ln (2/3), \\
I_t(x_t=1,y_t=1) &= \ln\frac{1/3}{(1/2)\cdot(1/2)} = \ln (4/3),
\end{aligned}
\ees
so that $\left<\Delta I\right> = (1/3)\ln (4/3) + (1/6)\ln (2/3) + (1/6)\ln (2/3) + (1/3)\ln (4/3) \approx \ln (1.06)$.
Thus by \eqreff{eq:coro} we obtain \textcolor{black}{the lower bound of the average entropy production for the coupling that has been established until time $t$ from the uncorrelated initial state, as follows: $\left<\sigma \right> \ge \left<\Delta I\right> \approx \ln 1.06.$
If there is no measurement error at final time $\tau$ such that $p_{\tau}(x_\tau=0,y_\tau=1)=p_{\tau}(x_\tau=1,y_\tau=0)=0$ and  $p_{\tau}(x_\tau=0,y_\tau=0)=p_{\tau}(x_\tau=1,y_\tau=1)=1/2$, then we may have $\left<\sigma \right>\ge \left<\Delta I\right> = \ln 2$, which is greater than $\ln 1.06$.}


\subsection{Feedback Control}
\textcolor{black}{
Unlike the case in \cite{sagawa2}, we need not to exchange subsystems $X$ and $Y$ to consider feedback control after the measurement. Thus we proceed continuously to feedback control immediately after each measurement process at time $\tau$ (see Figure 2b). We assume that correlation $I_\tau(x_\tau,y_\tau)$ at time $\tau$ is given by the values in \eqreff{eq:I0} and final correlation at later time $\tau'$ is zero, i.e., $I_{\tau'}(x_{\tau'},y_{\tau'})=0$. By feedback control, we mean that external parameter $\lambda_t$ for $\tau\le t\le \tau'$ is manipulated in a pre-determined manner \cite{sagawa2}, while systems $X$ and $Y$ co-evolve in time, 
\textcolor{black}{such that the established correlation is used as a source of work while $I_t(x_t,y_t)$ for $\tau \le t\le \tau'$ is decreased, not necessarily monotonically. \eqreff{eq:main2} provides an exact relation on the energetics of this process.} We rewrite its corollary, \eqreff{eq:coro2}, with respect to extractable work $W_{\rm ext}:=-W$ as follows: 
\be\label{eq:coro3}
\left< W_{\rm ext}\right> \le -\left< \Delta{\cal F}_x + \Delta{\cal F}_y + \frac{\Delta I}{\beta} \right>.
\ee
Then the extractable work on top of the conventional bound, $-\left< \Delta{\cal F}_x + \Delta{\cal F}_y\right>$, is additionally given by $-\Delta I / \beta =\ln (1.06)$,} \textcolor{black}{which comes from the consumption of the established correlation.}


\section{Conclusions}

\textcolor{black}{
We have proved the fluctuation theorem of information exchange, \eqreff{eq:main}, which holds even during the co-evolution of two systems that exchange information with each other. \eqreff{eq:main} tells us that establishing correlation between two systems necessarily accompanies entropy production which is contributed by both systems and the heat reservoir, as expressed in \eqreff{eq:sigma} and \eqreff{eq:s}. We have also proved, as a corollary of \eqreff{eq:main}, the fluctuation theorem of information exchange with energetic terms, \eqreff{eq:main2}, under the assumption of weak coupling between the two subsystems. \eqreff{eq:main2} reveals the exact relationship between non-equilibrium free energy of both sub-systems and mutual information that is established/consumed through their interactions. This more generalized framework than that in \cite{sagawa2}, enables us to apply thermodynamics of information to biological systems, where molecules generate/consume correlations through their information processing mechanisms \cite{Becker2015, Thomas2017, Ouldridge2017}. 
\textcolor{black}{Since the new framework is applicable to fully non-equilibrium situations, thermodynamic coupling during a dynamic allosteric transition, for example, may be analyzed based on this theoretical framework beyond current equilibrium thermodynamic approach \cite{tsai2014unified, cuendet2016allostery}.}
}





\noindent
{Acknowledgments.}
L.J. was supported by the National Research Foundation of Korea grant funded by the Korean Government (NRF-2010-0006733, NRF-2012R1A1A2042932, NRF-2016R1D1A1B02011106), and in part by 
Kwangwoon University Research Grant in 2016.

\bibliography{bib}

\begin{thebibliography}{37}%
\makeatletter
\providecommand \@ifxundefined [1]{%
 \@ifx{#1\undefined}
}%
\providecommand \@ifnum [1]{%
 \ifnum #1\expandafter \@firstoftwo
 \else \expandafter \@secondoftwo
 \fi
}%
\providecommand \@ifx [1]{%
 \ifx #1\expandafter \@firstoftwo
 \else \expandafter \@secondoftwo
 \fi
}%
\providecommand \natexlab [1]{#1}%
\providecommand \enquote  [1]{``#1''}%
\providecommand \bibnamefont  [1]{#1}%
\providecommand \bibfnamefont [1]{#1}%
\providecommand \citenamefont [1]{#1}%
\providecommand \href@noop [0]{\@secondoftwo}%
\providecommand \href [0]{\begingroup \@sanitize@url \@href}%
\providecommand \@href[1]{\@@startlink{#1}\@@href}%
\providecommand \@@href[1]{\endgroup#1\@@endlink}%
\providecommand \@sanitize@url [0]{\catcode `\\12\catcode `\$12\catcode
  `\&12\catcode `\#12\catcode `\^12\catcode `\_12\catcode `\%12\relax}%
\providecommand \@@startlink[1]{}%
\providecommand \@@endlink[0]{}%
\providecommand \url  [0]{\begingroup\@sanitize@url \@url }%
\providecommand \@url [1]{\endgroup\@href {#1}{\urlprefix }}%
\providecommand \urlprefix  [0]{URL }%
\providecommand \Eprint [0]{\href }%
\providecommand \doibase [0]{http://dx.doi.org/}%
\providecommand \selectlanguage [0]{\@gobble}%
\providecommand \bibinfo  [0]{\@secondoftwo}%
\providecommand \bibfield  [0]{\@secondoftwo}%
\providecommand \translation [1]{[#1]}%
\providecommand \BibitemOpen [0]{}%
\providecommand \bibitemStop [0]{}%
\providecommand \bibitemNoStop [0]{.\EOS\space}%
\providecommand \EOS [0]{\spacefactor3000\relax}%
\providecommand \BibitemShut  [1]{\csname bibitem#1\endcsname}%
\let\auto@bib@innerbib\@empty
\bibitem [{\citenamefont {Hartwell}\ \emph {et~al.}(1999)\citenamefont
  {Hartwell}, \citenamefont {Hopfield}, \citenamefont {Leibler},\ and\
  \citenamefont {Murray}}]{hartwell1999molecular}%
  \BibitemOpen
  \bibfield  {author} {\bibinfo {author} {\bibfnamefont {L.~H.}\ \bibnamefont
  {Hartwell}}, \bibinfo {author} {\bibfnamefont {J.~J.}\ \bibnamefont
  {Hopfield}}, \bibinfo {author} {\bibfnamefont {S.}~\bibnamefont {Leibler}}, \
  and\ \bibinfo {author} {\bibfnamefont {A.~W.}\ \bibnamefont {Murray}},\
  }\href@noop {} {\bibfield  {journal} {\bibinfo  {journal} {Nature}\ }\textbf
  {\bibinfo {volume} {402}},\ \bibinfo {pages} {C47} (\bibinfo {year}
  {1999})}\BibitemShut {NoStop}%
\bibitem [{\citenamefont {Crofts}(2007)}]{crofts2007life}%
  \BibitemOpen
  \bibfield  {author} {\bibinfo {author} {\bibfnamefont {A.~R.}\ \bibnamefont
  {Crofts}},\ }\href@noop {} {\bibfield  {journal} {\bibinfo  {journal}
  {Complexity}\ }\textbf {\bibinfo {volume} {13}},\ \bibinfo {pages} {14}
  (\bibinfo {year} {2007})}\BibitemShut {NoStop}%
\bibitem [{\citenamefont {Cheong}\ \emph {et~al.}(2011)\citenamefont {Cheong},
  \citenamefont {Rhee}, \citenamefont {Wang}, \citenamefont {Nemenman},\ and\
  \citenamefont {Levchenko}}]{cheong2011information}%
  \BibitemOpen
  \bibfield  {author} {\bibinfo {author} {\bibfnamefont {R.}~\bibnamefont
  {Cheong}}, \bibinfo {author} {\bibfnamefont {A.}~\bibnamefont {Rhee}},
  \bibinfo {author} {\bibfnamefont {C.~J.}\ \bibnamefont {Wang}}, \bibinfo
  {author} {\bibfnamefont {I.}~\bibnamefont {Nemenman}}, \ and\ \bibinfo
  {author} {\bibfnamefont {A.}~\bibnamefont {Levchenko}},\ }\href@noop {}
  {\bibfield  {journal} {\bibinfo  {journal} {science}\ }\textbf {\bibinfo
  {volume} {334}},\ \bibinfo {pages} {354} (\bibinfo {year}
  {2011})}\BibitemShut {NoStop}%
\bibitem [{\citenamefont {McGrath}\ \emph {et~al.}(2017)\citenamefont
  {McGrath}, \citenamefont {Jones}, \citenamefont {ten Wolde},\ and\
  \citenamefont {Ouldridge}}]{Thomas2017}%
  \BibitemOpen
  \bibfield  {author} {\bibinfo {author} {\bibfnamefont {T.}~\bibnamefont
  {McGrath}}, \bibinfo {author} {\bibfnamefont {N.~S.}\ \bibnamefont {Jones}},
  \bibinfo {author} {\bibfnamefont {P.~R.}\ \bibnamefont {ten Wolde}}, \ and\
  \bibinfo {author} {\bibfnamefont {T.~E.}\ \bibnamefont {Ouldridge}},\ }\href
  {\doibase 10.1103/PhysRevLett.118.028101} {\bibfield  {journal} {\bibinfo
  {journal} {Phys. Rev. Lett.}\ }\textbf {\bibinfo {volume} {118}},\ \bibinfo
  {pages} {028101} (\bibinfo {year} {2017})}\BibitemShut {NoStop}%
\bibitem [{\citenamefont {Ouldridge}\ \emph {et~al.}(2017)\citenamefont
  {Ouldridge}, \citenamefont {Govern},\ and\ \citenamefont {ten
  Wolde}}]{Ouldridge2017}%
  \BibitemOpen
  \bibfield  {author} {\bibinfo {author} {\bibfnamefont {T.~E.}\ \bibnamefont
  {Ouldridge}}, \bibinfo {author} {\bibfnamefont {C.~C.}\ \bibnamefont
  {Govern}}, \ and\ \bibinfo {author} {\bibfnamefont {P.~R.}\ \bibnamefont {ten
  Wolde}},\ }\href {\doibase 10.1103/PhysRevX.7.021004} {\bibfield  {journal}
  {\bibinfo  {journal} {Phys. Rev. X}\ }\textbf {\bibinfo {volume} {7}},\
  \bibinfo {pages} {021004} (\bibinfo {year} {2017})}\BibitemShut {NoStop}%
\bibitem [{\citenamefont {Becker}\ \emph {et~al.}(2015)\citenamefont {Becker},
  \citenamefont {Mugler},\ and\ \citenamefont {ten Wolde}}]{Becker2015}%
  \BibitemOpen
  \bibfield  {author} {\bibinfo {author} {\bibfnamefont {N.~B.}\ \bibnamefont
  {Becker}}, \bibinfo {author} {\bibfnamefont {A.}~\bibnamefont {Mugler}}, \
  and\ \bibinfo {author} {\bibfnamefont {P.~R.}\ \bibnamefont {ten Wolde}},\
  }\href {\doibase 10.1103/PhysRevLett.115.258103} {\bibfield  {journal}
  {\bibinfo  {journal} {Phys. Rev. Lett.}\ }\textbf {\bibinfo {volume} {115}},\
  \bibinfo {pages} {258103} (\bibinfo {year} {2015})}\BibitemShut {NoStop}%
\bibitem [{\citenamefont {Cheng}\ \emph {et~al.}(2016)\citenamefont {Cheng},
  \citenamefont {Liu}, \citenamefont {Shen},\ and\ \citenamefont
  {Zhao}}]{cheng2016}%
  \BibitemOpen
  \bibfield  {author} {\bibinfo {author} {\bibfnamefont {F.}~\bibnamefont
  {Cheng}}, \bibinfo {author} {\bibfnamefont {C.}~\bibnamefont {Liu}}, \bibinfo
  {author} {\bibfnamefont {B.}~\bibnamefont {Shen}}, \ and\ \bibinfo {author}
  {\bibfnamefont {Z.}~\bibnamefont {Zhao}},\ }\href@noop {} {\bibfield
  {journal} {\bibinfo  {journal} {BMC Systems Biology}\ }\textbf {\bibinfo
  {volume} {10}},\ \bibinfo {pages} {65} (\bibinfo {year} {2016})}\BibitemShut
  {NoStop}%
\bibitem [{\citenamefont {Whitsett}\ \emph {et~al.}(2016)\citenamefont
  {Whitsett}, \citenamefont {Guo}, \citenamefont {Xu}, \citenamefont {Bao},\
  and\ \citenamefont {Wagner}}]{Guo2016}%
  \BibitemOpen
  \bibfield  {author} {\bibinfo {author} {\bibfnamefont {J.~A.}\ \bibnamefont
  {Whitsett}}, \bibinfo {author} {\bibfnamefont {M.}~\bibnamefont {Guo}},
  \bibinfo {author} {\bibfnamefont {Y.}~\bibnamefont {Xu}}, \bibinfo {author}
  {\bibfnamefont {E.~L.}\ \bibnamefont {Bao}}, \ and\ \bibinfo {author}
  {\bibfnamefont {M.}~\bibnamefont {Wagner}},\ }\href@noop {} {\bibfield
  {journal} {\bibinfo  {journal} {Nucleic Acids Research}\ }\textbf {\bibinfo
  {volume} {45}},\ \bibinfo {pages} {e54} (\bibinfo {year} {2016})}\BibitemShut
  {NoStop}%
\bibitem [{oli(2018)}]{olimpio2018}%
  \BibitemOpen
  \href@noop {} {\bibfield  {journal} {\bibinfo  {journal} {iScience}\ }\textbf
  {\bibinfo {volume} {2}},\ \bibinfo {pages} {27 } (\bibinfo {year}
  {2018})}\BibitemShut {NoStop}%
\bibitem [{\citenamefont {Maire}\ and\ \citenamefont {Youk}(2015)}]{maire2015}%
  \BibitemOpen
  \bibfield  {author} {\bibinfo {author} {\bibfnamefont {T.}~\bibnamefont
  {Maire}}\ and\ \bibinfo {author} {\bibfnamefont {H.}~\bibnamefont {Youk}},\
  }\href@noop {} {\bibfield  {journal} {\bibinfo  {journal} {Cell Systems}\
  }\textbf {\bibinfo {volume} {1}},\ \bibinfo {pages} {349 } (\bibinfo {year}
  {2015})}\BibitemShut {NoStop}%
\bibitem [{\citenamefont {Mehta}\ and\ \citenamefont
  {Schwab}(2012)}]{mehta2012energetic}%
  \BibitemOpen
  \bibfield  {author} {\bibinfo {author} {\bibfnamefont {P.}~\bibnamefont
  {Mehta}}\ and\ \bibinfo {author} {\bibfnamefont {D.~J.}\ \bibnamefont
  {Schwab}},\ }\href@noop {} {\bibfield  {journal} {\bibinfo  {journal}
  {Proceedings of the National Academy of Sciences}\ }\textbf {\bibinfo
  {volume} {109}},\ \bibinfo {pages} {17978} (\bibinfo {year}
  {2012})}\BibitemShut {NoStop}%
\bibitem [{\citenamefont {Govern}\ and\ \citenamefont {ten
  Wolde}(2014)}]{govern2014energy}%
  \BibitemOpen
  \bibfield  {author} {\bibinfo {author} {\bibfnamefont {C.~C.}\ \bibnamefont
  {Govern}}\ and\ \bibinfo {author} {\bibfnamefont {P.~R.}\ \bibnamefont {ten
  Wolde}},\ }\href@noop {} {\bibfield  {journal} {\bibinfo  {journal} {Physical
  review letters}\ }\textbf {\bibinfo {volume} {113}},\ \bibinfo {pages}
  {258102} (\bibinfo {year} {2014})}\BibitemShut {NoStop}%
\bibitem [{\citenamefont {Leff}\ and\ \citenamefont {Rex}(2014)}]{leff2014}%
  \BibitemOpen
  \bibfield  {author} {\bibinfo {author} {\bibfnamefont {H.~S.}\ \bibnamefont
  {Leff}}\ and\ \bibinfo {author} {\bibfnamefont {A.~F.}\ \bibnamefont {Rex}},\
  }\href@noop {} {\emph {\bibinfo {title} {Maxwell's demon: entropy,
  information, computing}}}\ (\bibinfo  {publisher} {Princeton University
  Press},\ \bibinfo {year} {2014})\BibitemShut {NoStop}%
\bibitem [{\citenamefont {Landauer}\ \emph {et~al.}(1991)\citenamefont
  {Landauer} \emph {et~al.}}]{landauer1991information}%
  \BibitemOpen
  \bibfield  {author} {\bibinfo {author} {\bibfnamefont {R.}~\bibnamefont
  {Landauer}} \emph {et~al.},\ }\href@noop {} {\bibfield  {journal} {\bibinfo
  {journal} {Physics Today}\ }\textbf {\bibinfo {volume} {44}},\ \bibinfo
  {pages} {23} (\bibinfo {year} {1991})}\BibitemShut {NoStop}%
\bibitem [{\citenamefont {Szilard}(1964)}]{szilard}%
  \BibitemOpen
  \bibfield  {author} {\bibinfo {author} {\bibfnamefont {L.}~\bibnamefont
  {Szilard}},\ }\href@noop {} {\bibfield  {journal} {\bibinfo  {journal}
  {Behavioral Science}\ }\textbf {\bibinfo {volume} {9}},\ \bibinfo {pages}
  {301} (\bibinfo {year} {1964})}\BibitemShut {NoStop}%
\bibitem [{\citenamefont {Sagawa}\ and\ \citenamefont {Ueda}(2012)}]{sagawa2}%
  \BibitemOpen
  \bibfield  {author} {\bibinfo {author} {\bibfnamefont {T.}~\bibnamefont
  {Sagawa}}\ and\ \bibinfo {author} {\bibfnamefont {M.}~\bibnamefont {Ueda}},\
  }\href@noop {} {\bibfield  {journal} {\bibinfo  {journal} {Phys. Rev. Lett.}\
  }\textbf {\bibinfo {volume} {109}},\ \bibinfo {pages} {180602} (\bibinfo
  {year} {2012})}\BibitemShut {NoStop}%
\bibitem [{\citenamefont {Cover}\ and\ \citenamefont {Thomas}(2012)}]{cover}%
  \BibitemOpen
  \bibfield  {author} {\bibinfo {author} {\bibfnamefont {T.~M.}\ \bibnamefont
  {Cover}}\ and\ \bibinfo {author} {\bibfnamefont {J.~A.}\ \bibnamefont
  {Thomas}},\ }\href@noop {} {\emph {\bibinfo {title} {Elements of information
  theory}}}\ (\bibinfo  {publisher} {John Wiley \& Sons},\ \bibinfo {year}
  {2012})\BibitemShut {NoStop}%
\bibitem [{\citenamefont {Tsai}\ and\ \citenamefont
  {Nussinov}(2014)}]{tsai2014unified}%
  \BibitemOpen
  \bibfield  {author} {\bibinfo {author} {\bibfnamefont {C.-J.}\ \bibnamefont
  {Tsai}}\ and\ \bibinfo {author} {\bibfnamefont {R.}~\bibnamefont
  {Nussinov}},\ }\href@noop {} {\bibfield  {journal} {\bibinfo  {journal} {PLoS
  computational biology}\ }\textbf {\bibinfo {volume} {10}},\ \bibinfo {pages}
  {e1003394} (\bibinfo {year} {2014})}\BibitemShut {NoStop}%
\bibitem [{\citenamefont {Cuendet}\ \emph {et~al.}(2016)\citenamefont
  {Cuendet}, \citenamefont {Weinstein},\ and\ \citenamefont
  {LeVine}}]{cuendet2016allostery}%
  \BibitemOpen
  \bibfield  {author} {\bibinfo {author} {\bibfnamefont {M.~A.}\ \bibnamefont
  {Cuendet}}, \bibinfo {author} {\bibfnamefont {H.}~\bibnamefont {Weinstein}},
  \ and\ \bibinfo {author} {\bibfnamefont {M.~V.}\ \bibnamefont {LeVine}},\
  }\href@noop {} {\bibfield  {journal} {\bibinfo  {journal} {Journal of
  chemical theory and computation}\ }\textbf {\bibinfo {volume} {12}},\
  \bibinfo {pages} {5758} (\bibinfo {year} {2016})}\BibitemShut {NoStop}%
\bibitem [{\citenamefont {Jarzynski}(2011)}]{jarReview}%
  \BibitemOpen
  \bibfield  {author} {\bibinfo {author} {\bibfnamefont {C.}~\bibnamefont
  {Jarzynski}},\ }\href@noop {} {\bibfield  {journal} {\bibinfo  {journal}
  {Annu. Rev. Codens. Matter Phys.}\ }\textbf {\bibinfo {volume} {2}},\
  \bibinfo {pages} {329} (\bibinfo {year} {2011})}\BibitemShut {NoStop}%
\bibitem [{\citenamefont {Seifert}(2012)}]{revSeifert}%
  \BibitemOpen
  \bibfield  {author} {\bibinfo {author} {\bibfnamefont {U.}~\bibnamefont
  {Seifert}},\ }\href@noop {} {\bibfield  {journal} {\bibinfo  {journal} {Rep.
  Prog. Phys.}\ }\textbf {\bibinfo {volume} {75}},\ \bibinfo {pages} {126001}
  (\bibinfo {year} {2012})}\BibitemShut {NoStop}%
\bibitem [{\citenamefont {Spinney}\ and\ \citenamefont {Ford}(2013)}]{review}%
  \BibitemOpen
  \bibfield  {author} {\bibinfo {author} {\bibfnamefont {R.}~\bibnamefont
  {Spinney}}\ and\ \bibinfo {author} {\bibfnamefont {I.}~\bibnamefont {Ford}},\
  }\enquote {\bibinfo {title} {Fluctuation relations: A pedagogical
  overview},}\ in\ \href@noop {} {\emph {\bibinfo {booktitle} {Nonequilibrium
  Statistical Physics of Small Systems}}}\ (\bibinfo  {publisher} {Wiley-VCH
  Verlag GmbH \& Co. KGaA},\ \bibinfo {year} {2013})\ pp.\ \bibinfo {pages}
  {3--56}\BibitemShut {NoStop}%
\bibitem [{\citenamefont {Crooks}(1999)}]{crooks99}%
  \BibitemOpen
  \bibfield  {author} {\bibinfo {author} {\bibfnamefont {G.~E.}\ \bibnamefont
  {Crooks}},\ }\href@noop {} {\bibfield  {journal} {\bibinfo  {journal} {Phys.
  Rev. E}\ }\textbf {\bibinfo {volume} {60}},\ \bibinfo {pages} {2721}
  (\bibinfo {year} {1999})}\BibitemShut {NoStop}%
\bibitem [{\citenamefont {Seifert}(2005)}]{seifert05}%
  \BibitemOpen
  \bibfield  {author} {\bibinfo {author} {\bibfnamefont {U.}~\bibnamefont
  {Seifert}},\ }\href@noop {} {\bibfield  {journal} {\bibinfo  {journal} {Phys.
  Rev. Lett.}\ }\textbf {\bibinfo {volume} {95}},\ \bibinfo {pages} {040602}
  (\bibinfo {year} {2005})}\BibitemShut {NoStop}%
\bibitem [{\citenamefont {Ponmurugan}(2010)}]{ponmurugan2010}%
  \BibitemOpen
  \bibfield  {author} {\bibinfo {author} {\bibfnamefont {M.}~\bibnamefont
  {Ponmurugan}},\ }\href@noop {} {\bibfield  {journal} {\bibinfo  {journal}
  {Physical Review E}\ }\textbf {\bibinfo {volume} {82}},\ \bibinfo {pages}
  {031129} (\bibinfo {year} {2010})}\BibitemShut {NoStop}%
\bibitem [{\citenamefont {Horowitz}\ and\ \citenamefont
  {Vaikuntanathan}(2010)}]{horowitz2010}%
  \BibitemOpen
  \bibfield  {author} {\bibinfo {author} {\bibfnamefont {J.~M.}\ \bibnamefont
  {Horowitz}}\ and\ \bibinfo {author} {\bibfnamefont {S.}~\bibnamefont
  {Vaikuntanathan}},\ }\href@noop {} {\bibfield  {journal} {\bibinfo  {journal}
  {Physical Review E}\ }\textbf {\bibinfo {volume} {82}},\ \bibinfo {pages}
  {061120} (\bibinfo {year} {2010})}\BibitemShut {NoStop}%
\bibitem [{\citenamefont {Kurchan}(1998)}]{kur}%
  \BibitemOpen
  \bibfield  {author} {\bibinfo {author} {\bibfnamefont {J.}~\bibnamefont
  {Kurchan}},\ }\href@noop {} {\bibfield  {journal} {\bibinfo  {journal}
  {Journal of Physics A: Mathematical and General}\ }\textbf {\bibinfo {volume}
  {31}},\ \bibinfo {pages} {3719} (\bibinfo {year} {1998})}\BibitemShut
  {NoStop}%
\bibitem [{\citenamefont {Maes}(1999)}]{maes1999}%
  \BibitemOpen
  \bibfield  {author} {\bibinfo {author} {\bibfnamefont {C.}~\bibnamefont
  {Maes}},\ }\href@noop {} {\bibfield  {journal} {\bibinfo  {journal} {Journal
  of statistical physics}\ }\textbf {\bibinfo {volume} {95}},\ \bibinfo {pages}
  {367} (\bibinfo {year} {1999})}\BibitemShut {NoStop}%
\bibitem [{\citenamefont {Jarzynski}(2000)}]{jar2000}%
  \BibitemOpen
  \bibfield  {author} {\bibinfo {author} {\bibfnamefont {C.}~\bibnamefont
  {Jarzynski}},\ }\href@noop {} {\bibfield  {journal} {\bibinfo  {journal} {J.
  Stat. Phys.}\ }\textbf {\bibinfo {volume} {98}},\ \bibinfo {pages} {77}
  (\bibinfo {year} {2000})}\BibitemShut {NoStop}%
\bibitem [{\citenamefont {Goldstein}\ \emph {et~al.}(2001)\citenamefont
  {Goldstein}, \citenamefont {Jr.},\ and\ \citenamefont {Safko}}]{goldstein}%
  \BibitemOpen
  \bibfield  {author} {\bibinfo {author} {\bibfnamefont {H.}~\bibnamefont
  {Goldstein}}, \bibinfo {author} {\bibfnamefont {C.~P.~P.}\ \bibnamefont
  {Jr.}}, \ and\ \bibinfo {author} {\bibfnamefont {J.~L.}\ \bibnamefont
  {Safko}},\ }\href@noop {} {\emph {\bibinfo {title} {Classical Mechanics (3rd
  Edition)}}}\ (\bibinfo  {publisher} {Pearson},\ \bibinfo {year}
  {2001})\BibitemShut {NoStop}%
\bibitem [{\citenamefont {Parrondo}\ \emph {et~al.}(2015)\citenamefont
  {Parrondo}, \citenamefont {Horowitz},\ and\ \citenamefont
  {Sagawa}}]{parrondo2015}%
  \BibitemOpen
  \bibfield  {author} {\bibinfo {author} {\bibfnamefont {J.~M.}\ \bibnamefont
  {Parrondo}}, \bibinfo {author} {\bibfnamefont {J.~M.}\ \bibnamefont
  {Horowitz}}, \ and\ \bibinfo {author} {\bibfnamefont {T.}~\bibnamefont
  {Sagawa}},\ }\href@noop {} {\bibfield  {journal} {\bibinfo  {journal} {Nature
  physics}\ }\textbf {\bibinfo {volume} {11}},\ \bibinfo {pages} {131}
  (\bibinfo {year} {2015})}\BibitemShut {NoStop}%
\bibitem [{\citenamefont {Kawai}\ \emph {et~al.}(2007)\citenamefont {Kawai},
  \citenamefont {Parrondo},\ and\ \citenamefont {den Broeck}}]{kawai}%
  \BibitemOpen
  \bibfield  {author} {\bibinfo {author} {\bibfnamefont {R.}~\bibnamefont
  {Kawai}}, \bibinfo {author} {\bibfnamefont {J.~M.~R.}\ \bibnamefont
  {Parrondo}}, \ and\ \bibinfo {author} {\bibfnamefont {C.~V.}\ \bibnamefont
  {den Broeck}},\ }\href@noop {} {\bibfield  {journal} {\bibinfo  {journal}
  {Phys. Rev. Lett.}\ }\textbf {\bibinfo {volume} {98}},\ \bibinfo {pages}
  {080602} (\bibinfo {year} {2007})}\BibitemShut {NoStop}%
\bibitem [{tak(2010)}]{takara}%
  \BibitemOpen
  \href@noop {} {\bibfield  {journal} {\bibinfo  {journal} {Physics Letters A}\
  }\textbf {\bibinfo {volume} {375}},\ \bibinfo {pages} {88 } (\bibinfo {year}
  {2010})}\BibitemShut {NoStop}%
\bibitem [{has(2010)}]{hasegawa}%
  \BibitemOpen
  \href@noop {} {\bibfield  {journal} {\bibinfo  {journal} {Physics Letters A}\
  }\textbf {\bibinfo {volume} {374}},\ \bibinfo {pages} {1001 } (\bibinfo
  {year} {2010})}\BibitemShut {NoStop}%
\bibitem [{\citenamefont {Esposito}\ and\ \citenamefont {Van~den
  Broeck}(2011)}]{esposito2011}%
  \BibitemOpen
  \bibfield  {author} {\bibinfo {author} {\bibfnamefont {M.}~\bibnamefont
  {Esposito}}\ and\ \bibinfo {author} {\bibfnamefont {C.}~\bibnamefont {Van~den
  Broeck}},\ }\href@noop {} {\bibfield  {journal} {\bibinfo  {journal}
  {Europhys. Lett.}\ }\textbf {\bibinfo {volume} {95}},\ \bibinfo {pages}
  {40004} (\bibinfo {year} {2011})}\BibitemShut {NoStop}%
\bibitem [{\citenamefont {Sagawa}\ and\ \citenamefont {Ueda}(2010)}]{sagawa}%
  \BibitemOpen
  \bibfield  {author} {\bibinfo {author} {\bibfnamefont {T.}~\bibnamefont
  {Sagawa}}\ and\ \bibinfo {author} {\bibfnamefont {M.}~\bibnamefont {Ueda}},\
  }\href@noop {} {\bibfield  {journal} {\bibinfo  {journal} {Phys. Rev. Lett.}\
  }\textbf {\bibinfo {volume} {104}},\ \bibinfo {pages} {090602} (\bibinfo
  {year} {2010})}\BibitemShut {NoStop}%
\bibitem [{\citenamefont {Horowitz}\ and\ \citenamefont
  {Parrondo}(2011)}]{horowitz2011thermodynamic}%
  \BibitemOpen
  \bibfield  {author} {\bibinfo {author} {\bibfnamefont {J.~M.}\ \bibnamefont
  {Horowitz}}\ and\ \bibinfo {author} {\bibfnamefont {J.~M.}\ \bibnamefont
  {Parrondo}},\ }\href@noop {} {\bibfield  {journal} {\bibinfo  {journal} {EPL
  (Europhysics Letters)}\ }\textbf {\bibinfo {volume} {95}},\ \bibinfo {pages}
  {10005} (\bibinfo {year} {2011})}\BibitemShut {NoStop}%
\end{thebibliography}%

\end{document}